# Ultrahigh-pressure magnesium hydrosilicates as reservoirs of water in early Earth

Han-Fei Li,[1] Artem R. Oganov,[2] Haixu Cui,[3] Xiang-Feng Zhou,[4] Xiao Dong,[1,*] and Hui-Tian Wang[5]

[1] *Key Laboratory of Weak-Light Nonlinear Photonics and School of Physics, Nankai University, Tianjin 300071, China*
[2] *Skolkovo Institute of Science and Technology, Skolkovo Innovation Center, 30 bldg. 1 Bolshoy blvd., Moscow 121205, Russia*
[3] *College of Physics and Materials Science, Tianjin Normal University, Tianjin 300387, China*
[4] *Center for High Pressure Science, State Key Laboratory of Metastable Materials Science and Technology, School of Science, Yanshan University, Qinhuangdao 066004, China*
[5] *Collaborative Innovation Center of Advanced Microstructures, Nanjing University, Nanjing, 210093, China*
*\*xiao.dong@nankai.edu.cn*

**Abstract**

The origin of water on the Earth is a long-standing mystery, requiring a comprehensive search for hydrous compounds, stable at conditions of the deep Earth and made of Earth-abundant elements. Previous studies usually focused on the current geothermal situation in the Earth's mantle and ignored a possible difference in the past, such as the stage of the core–mantle separation. Here, using ab initio evolutionary structure prediction, we find that only two magnesium hydrosilicate phases are stable at megabar pressures, α-$Mg_2SiO_5H_2$ and β-$Mg_2SiO_5H_2$, stable at 262–338 GPa and >338 GPa, respectively (all these pressures now lie within the Earth's iron core). Both are superionic conductors with quasi-one-dimensional proton diffusion at relevant conditions. In the first 50-100 million years of Earth's history, before the Earth's core was formed, these must have existed in the Earth, hosting much of Earth's water. As dense iron alloys segregated to form the Earth's core, $Mg_2SiO_5H_2$ phases decomposed and released water. Thus, now-extinct $Mg_2SiO_5H_2$ phases have likely contributed in a major way to the evolution of our planet.

"Where did Earth's water come from" is a long-standing mystery, essential for understanding how life appeared and how the dynamics of Earth's interiors evolved with time. Currently, there are two conflicting views on this matter: (1) that water is primordial, i.e. that the Earth had acquired most of its water during accretion, or (2) that water had been donated later by water-rich aerolites. The first hypothesis essentially means that water is released from inside the Earth ("hell"), whereas the second one states that water comes from the outer space ("heaven"). Recently, increasing evidence supported the first hypothesis [1-3]. The deuterium/hydrogen (D/H) ratio, considered as the fingerprint of the origin of water, offers a persuasive argument: the Earth's deep mantle has a low D/H ratio quite close to that of enstatite chondrite meteorites [4], which are the fundamental building blocks of the young Earth, indicating that water within the Earth's interior may have come directly from the protosolar nebula [1].

However, this hypothesis raises several questions. Compared with other planetary materials such as iron and silicates, water has a much lower condensation temperature and therefore would have been released to space at the high surface temperature of the newborn Earth and then by the Moon-forming impact. To avoid complete loss, water must have been stored inside the Earth in the planet's neonatal accretion period. Hydrous minerals, such as hydrous silicates, are prime candidates for such reservoir.

Several hydrous magnesium silicates have been reported in the previous study [5] – which is natural, given that Mg, Si and O are the most abundant elements in the Earth's mantle. Much scientific interest was attracted to phases B and C ($Mg_{10}Si_3O_{18}H_4$), stable up to ~17.8 GPa [6], phases D, F, and G ($Mg_{1.14}Si_{1.73}H_{2.81}O_6$), stable at least up to 22.5 GPa and around 1000 °C [6,7]. These hydrous phases are probably the essential water reservoirs in the upper mantle and transition zone. Recently discovered phase H ($MgSiO_4H_2$) stable at high pressures above 48 GPa, may preserve water in the lower mantle [8], however, later it was shown to dissociate into $MgSiO_3$ (bridgmanite) and $H_2O$ (ice-VIII) when pressure is further increased to ~52 GPa (first-principles prediction) [9] or to ~60 GPa (experiment) [10]. Furthermore, phase H is unstable at high temperatures, its upper dissociation boundary being ~1500 K [11] predicted using first-principles calculations within the quasiharmonic approximation (QHA). Such temperatures can only be found in the coldest parts of the lower mantle, in particular, in subducted lithospheric slabs, which are typically ~500 K colder than normal mantle, thus restricting possible abundance of phase H in the mantle. Further work [8] showed that Al-rich phase H can survive in a much wider P-T field on the phase diagram. Other hydrous phases explored to date are hydrous aluminum silicates [12-14] and iron hydroxides [15,16]. Note that so far research focused on the hydrous minerals of today's mantle, where pressures are below 136 GPa – higher pressures are only found in the Earth's core, which is made of iron alloys and has no silicates or hydroxides. But this is so today – our report draws attention to the times before the formation of the core, when silicates existed throughout the Earth, at much higher pressures.

The core–mantle separation [17] is the most significant process in the early history of the Earth. It is believed that protoplanets had been formed as a result of collisions of smaller bodies (planetesimals), which had previously condensed from solid debris present in the original nebula. Planetesimals contained iron and silicates either already differentiated or mixed together. With time, dense iron alloys settled to the center of the Earth, forming its core. Complex geochemical processes that accompanied core formation have significantly affected the chemistry and evolution of the Earth. Previous studies [17] of the core–mantle separation mainly focused on transition metals, especially siderophile ("iron-loving") ones, but there is more to it. In the young Earth, before the core–mantle separation, silicates extended much deeper and have been exposed to much higher pressures than now. Here we show that hitherto unknown hydrous phases are stable at such conditions – they should have existed at times before the formation of the core. These ephemeral, now extinct, mineral phases were probably a powerful factor affecting the fate of our planet.

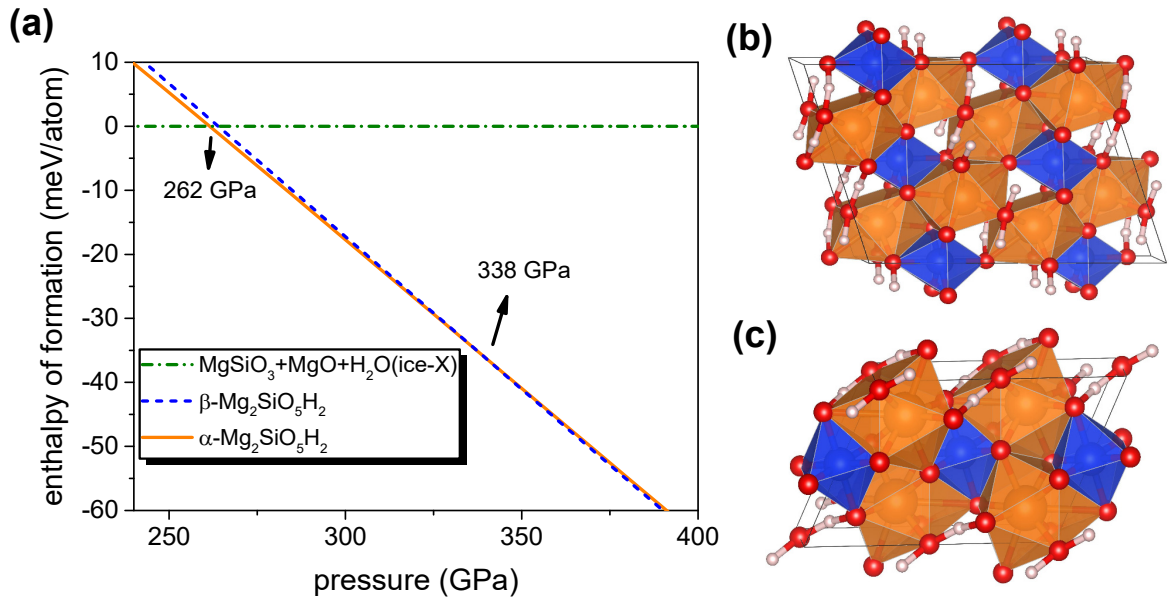

Fig. 1. Stability and structures of $Mg_2SiO_5H_2$ phases. (a) Enthalpy of formation of α-, β-$Mg_2SiO_5H_2$, relative to the enthalpy of $MgSiO_3$ + $MgO$ + $H_2O$ assemblage, as a function of the pressure. Note that zero point energy (ZPE) is not included in these enthalpies. (b) Crystal structure of α-$Mg_2SiO_5H_2$ in the [010] direction. (c) Crystal structure of β-$Mg_2SiO_5H_2$ in the [010] direction. The Mg, Si, O, and H atoms are shown as orange, blue, red, and white spheres, respectively.

The ternary system $MgO$–$SiO_2$–$H_2O$ was studied using a variable-composition evolutionary structure prediction algorithm as implemented in the USPEX code [18-20]. Structure relaxations and enthalpy calculations were done using the Perdew–Burke–Ernzerhof (PBE) functional [21] in the framework of the all-electron projector augmented wave (PAW) method [22] as implemented in the VASP code [23]. The PAW potentials used for Mg, Si, O, and H atoms treated $2s^22p^63s^2$, $3s^23p^2$, $2s^22p^4$, and $1s^1$ electrons as valence, and had core radii of 1.7, 1.9, 1.1, and 0.8 a.u., respectively. We used a plane-wave kinetic energy cutoff of 1200 eV and the Brillouin zone sampling with a resolution of $2\pi \times 0.03$ Å$^{-1}$, which showed excellent convergence of the energy differences, stress tensors, and structural parameters. Variable-composition structure searches were performed at pressures of 50, 100, 200, 300, 400, 500, and 1000 GPa, allowing up to 50 atoms per primitive cell. We also explored the effects of temperature on stability using the QHA, for which the phonon calculations were performed for all relevant structures using the PHONOPY code [24]. For each structure, phonons were computed at 30 different volumes to predict the Gibbs free energy.

Our comprehensive search revealed that only $MgSiO_3·MgO·H_2O$ – which we write as $Mg_2SiO_5H_2$ – has lower enthalpy than the mixture of $MgSiO_3$, $MgO$, and $H_2O$, or any other mixture at megabar pressures (Fig. 1). We found two thermodynamically stable modifications for this compound, as α-$Mg_2SiO_5H_2$ and β-$Mg_2SiO_5H_2$ with base-centered monoclinic lattices. Their transition pressure is 338 GPa. Below 262 GPa, dissociation to $MgSiO_3$, $MgO$, and $H_2O$ is favorable, whereas α-$Mg_2SiO_5H_2$ is thermodynamically stable at pressures of 262–338 GPa and β-$Mg_2SiO_5H_2$ has the lowest enthalpy at pressures >338 GPa.

The ternary phase diagram and convex hull based on enthalpies of formation at 200 and 300 GPa are shown as Fig. S1 in the Supplementary Materials. Counterintuitively, the formation of $Mg_2SiO_5H_2$ needs not much water at 300 GPa, but excess of magnesium with Mg/Si > 1 is necessary for its stability (Fig. S1b). Since there is excessive MgO in current lower mantle, we have reason to believe that the condition of excess of magnesium is satisfied.

For α-$Mg_2SiO_5H_2$ with space group $C2/c$ at 300 GPa, the relaxed lattice parameters are: $a$ = 10.19 Å, $b$ = 2.35 Å, $c$ = 7.49 Å, $\alpha = \gamma = 90°$, and $\beta = 108.79°$. The Mg atoms occupy the Wyckoff position 8$f$ (0.105, 0.742, 0.177), Si atoms — 4$c$ (0.25, 0.75, 0.5), H atoms — 8$f$ (0.055, 0.088, 0.386), and O atoms occupy positions 8$f$ (0.219, 0.246, 0.346), 8$f$ (0.407, 0.256, 0.004), and 4$e$ (0, 0.264, 0.25). Every Si atom is octahedrally coordinated by six O atoms, and the structure contains chains of corner-sharing $SiO_6$ octahedra. Mg atoms have eightfold (bicapped trigonal prismatic) coordination. Interestingly, every H atom is coordinated by two O atoms, forming symmetric hydrogen bonds, and occupies the longest edge of the Mg-centered prism. The Bravais cell contains four formula units of $Mg_2SiO_5H_2$ (40 atoms).

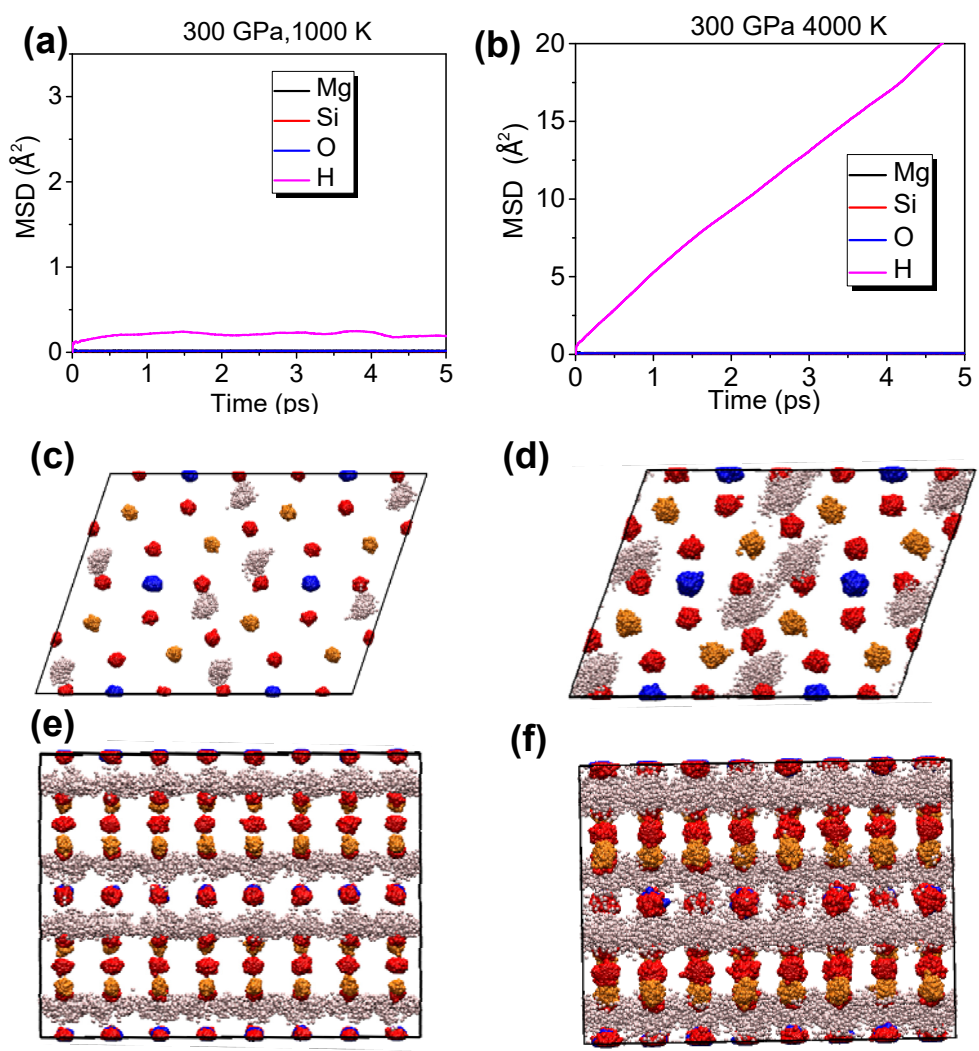

Fig. 2. Proton dynamics in $Mg_2SiO_5H_2$. (a, b) Mean square displacements of protons in α-$Mg_2SiO_5H_2$ at 300 GPa and (a) 1000 K, (b) 4000 K. (c–f) Projection of the atomic trajectories along the (c, d) [010] and (e, f) [100] direction of α-$Mg_2SiO_5H_2$ from the last 3 ps run representing (c, e) the normal (1000 K) and (d, f) the superionic (4000 K) states.

β-$Mg_2SiO_5H_2$, stable at pressures above 338 GPa, has space group $C2$. At 400 GPa, its lattice parameters are: $a$ = 7.29 Å, $b$ = 2.30 Å, $c$ = 5.12 Å, $\alpha = \gamma = 90°$, and $\beta = 113.70°$. The Mg atoms occupy the Wyckoff position $4c$ (0.323, 0.480, 0.791), Si atoms — $2b$ (0, 0.989, 0.5), H atoms — $4c$ (0.417, 0.133, 0.111), and O atoms occupy positions $4c$ (0.346, 0.483, 0.183), $4c$ (0.124, 0.486, 0.437), and $2a$ (0, 0.462, 0). Similar to α-$Mg_2SiO_5H_2$, each Si atom is octahedrally coordinated by six O atoms, whereas each Mg atom is nine-coordinate (capped square antiprismatic coordination by O atoms). Each H atom is coordinated by two O atoms, forming symmetric hydrogen bonds; hydrogens occupy two long edges of the Mg-centered antiprism. The Bravais cell contains two formula units of $Mg_2SiO_5H_2$ (20 atoms).

Geometrically, β-$Mg_2SiO_5H_2$ is a denser phase with a higher coordination number of Mg atoms. Nine-coordinate magnesium ions have not been seen before in Earth-forming minerals. The highest known coordination number of magnesium in silicates was 8 (in $MgSiO_3$ bridgmanite and post-perovskite). Even in the predicted ultrahigh-pressure phase of $Mg_2SiO_4$, expected to be stable at much higher pressures from 0.51 to 2.3 TPa, the coordination of Mg atoms is only eightfold [25]. The formation of β-$Mg_2SiO_5H_2$ indicates that hydration can increase the coordination of Mg, inducing a denser phase with a lower $PV$ term that stabilizes it at high pressure. At 300 GPa, both α-$Mg_2SiO_5H_2$ and β-$Mg_2SiO_5H_2$ are predicted to have very high densities of 6.203 and 6.207 g/cm³, respectively. Furthermore, $Mg_2SiO_5H_2$ contains 11.4 wt % of water, comparable to ~15 wt % in δ-$AlO_2H$ and phase H ($MgSiO_4H_2$), and higher than in most other reported hydrous silicates and hydroxides.

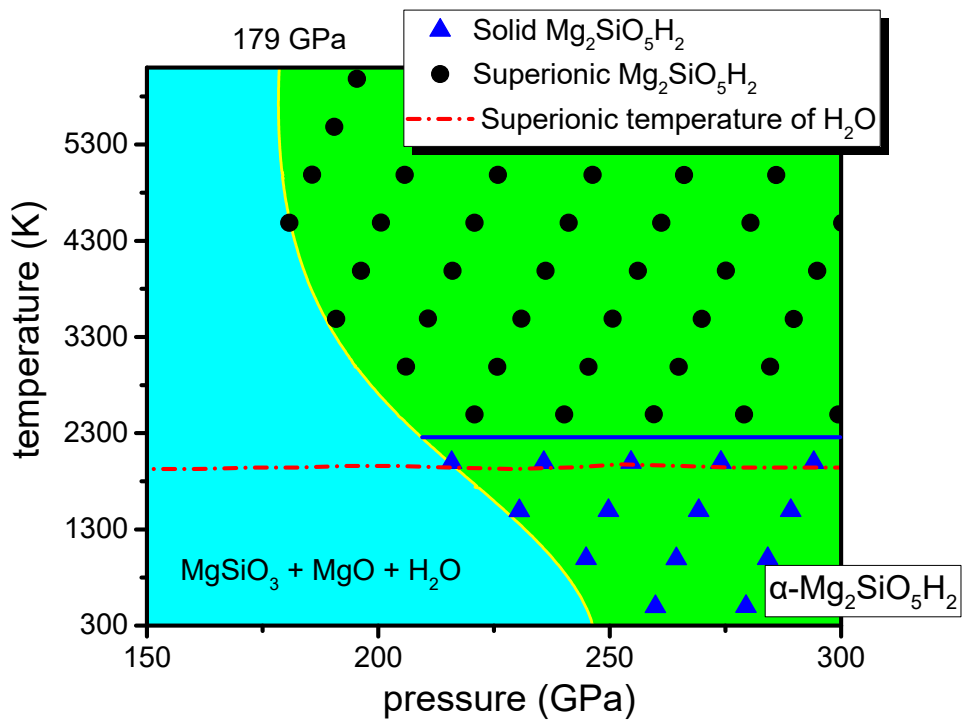

Fig. 3. Pressure–temperature stability field of α-$Mg_2SiO_5H_2$. The high-temperature dissociation boundary of α-$Mg_2SiO_5H_2$ is shown by the solid line. Note that with ZPE correction, the formation pressure at zero temperature shifts a little to 247 GPa. Normal and superionic states are marked by triangles and circles, respectively; their boundary at around 2300 K is indicated by a blue solid line. For comparison, the temperature of the superionic transition of $H_2O$ is shown by a dash-dotted line, based on the previous work [26].

To study the effect of temperature on $Mg_2SiO_5H_2$, we calculated the high-temperature dissociation phase boundary between $\alpha$-$Mg_2SiO_5H_2$ and $MgSiO_3 + MgO + H_2O$ using Gibbs free energies calculated within the QHA (Fig. 3). The field of stability of $\alpha$-$Mg_2SiO_5H_2$ widens as temperature rises, which is unexpected and different from phase H ($MgSiO_4H_2$) and other hydrous silicates: at 6000 K, $\alpha$-$Mg_2SiO_5H_2$ is stable already at 179 GPa. Our calculations show that this compound does not decompose at temperatures up to 8000 K at pressures of 200–400 GPa.

Like most hydrides at high pressures and temperatures, $Mg_2SiO_5H_2$ exhibits proton diffusion at high temperature, as we see from ab initio molecular dynamics (AIMD) simulations and the analysis of mean square displacements (MSDs) and atomic trajectories. At 300 GPa and 1000 K, the oscillations of the H, Mg, Si, and O atoms with respect to their equilibrium positions indicate that $Mg_2SiO_5H_2$ stays in the normal phase (Fig. 2a,c,e), whereas at 4000 K, the H atoms show fast diffusion on the picosecond timescale (Fig. 2b,d,f). This diffusion has quasi-one-dimensional character: proton diffusion favors the [010] direction.

The superionic transition temperature of $Mg_2SiO_5H_2$ (~2300 K) is nearly pressure-independent (Fig. 3), a situation similar to ice. AIMD simulations show that in the pressure range of thermodynamic stability of $\alpha$-$Mg_2SiO_5H_2$, its superionic state exists at temperatures above ~2500 K and does not melt spontaneously even at 8000 K and 250–400 GPa in our ab initio molecular dynamic simulation.

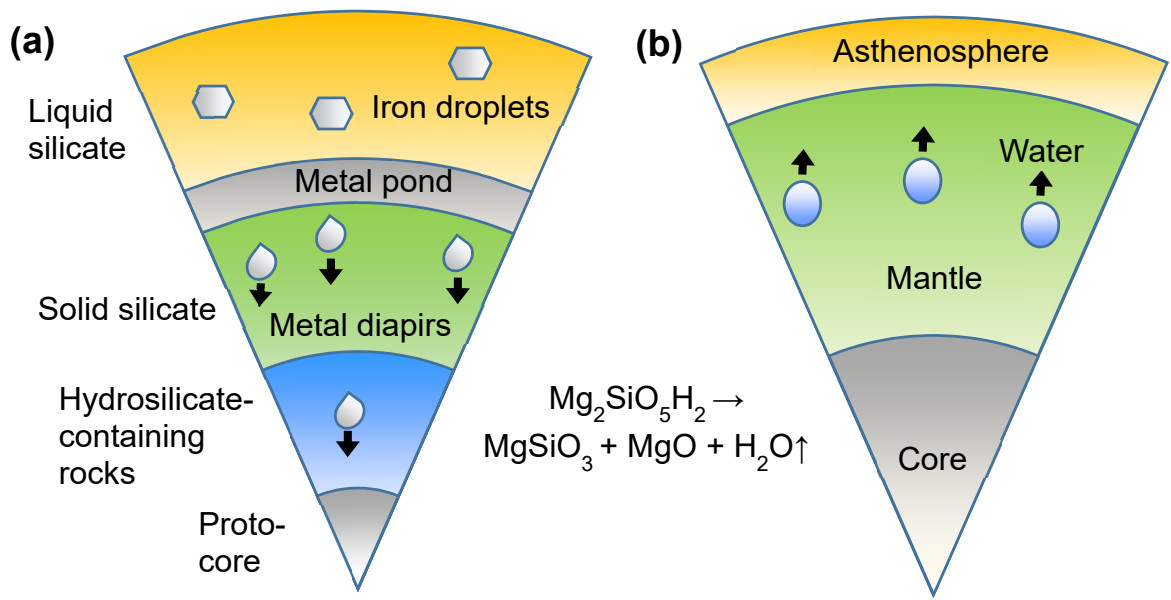

Fig. 4. Water storage model in (a) early stages of core formation, (b) after the core–mantle separation.

Before (and in also in the early stages of) the core-mantle separation, when silicates existed throughout the Earth, including depths with pressures well above 263 GPa, $Mg_2SiO_5H_2$ likely existed at those depths (Fig. 4). According to $^{182}Hf/^{182}W$ isotope chronometry, the core formation was essentially complete 62 ±10 million years after the iron meteorites were formed [27]. In this period, water could be stored in the form of $Mg_2SiO_5H_2$ - hidden in the deep interior, it avoided evaporation that inevitably occurred to water in near-surface regions, especially at the time of the Moon-forming impact. In an ideal situation, $Mg_2SiO_5H_2$ can hold water with nearly 8 times of current ocean mass, which is twice the entire water content of today's Earth (see Supplementary Materials which include Refs. [5,28,29]).

As the core grew, silicates were gradually displaced to shallower depths with lower pressures – and before the core reached its present size, $Mg_2SiO_5H_2$ was moved to regions where it is no longer stable, but had to dissociate: $Mg_2SiO_5H_2 \rightarrow MgSiO_3 + MgO + H_2O$. The released water would be gradually transported to the surface of the Earth, to form its hydrosphere. As to the other products of dissociation of $Mg_2SiO_5H_2$, namely $MgSiO_3$ and $MgO$, they are still in the lower mantle, playing the role of its main phases. Since transition zone and part of lower mantle are expected to have some ability to contain much water [5], current mantle-derived water and deep water cycle in mantle would begin at the time of $Mg_2SiO_5H_2$ dissociation. Although most ancient hydrous minerals would transport upwards and lose water by mantle convection in billions of years, perhaps some of the ancient relics survived from mantle convection and hided in the corners, waiting for the discovery of humankind.

In addition to creating the hydrosphere, the process of water release during the formation of the core could play several other important roles. First, some of the water could be dissolved in hot liquid iron core (and one can expect that as the core cools down, water solubility in it decreases, and water can still be gradually released from the core). Second, water could oxidize droplets of metallic iron in the lower mantle, to form $FeO_2H_x$ and free hydrogen (which will be degassed into the atmosphere and then lost to outer space, enriching the mantle in oxidized iron) – indeed, $FeO_2H_x$ was produced from water and iron at 86 GPa and 2200 K [16,30,31], and this compound was recently proposed to explain the observed ultralow-velocity zones at the core–mantle boundary [31]. The enrichment of the mantle in oxidized iron has likely contributed to the Great Oxidation Event, i.e. the emergence of free oxygen in the hitherto oxygen-free reducing atmosphere ~2.3 billion years ago. Third, the release of fluids (water and possibly hydrogen) would rheologically weaken mantle rocks, thereby enhancing their solid-state convection; such fluids also lower the melting temperature of the rocks, helping the formation of magmas [32].

Our hypothesis can explain why Mars is very dry compared to the Earth: because of its small size, Mars did not have sufficient pressures to form $Mg_2SiO_5H_2$ in its early history and thus had no ability to preserve water. Currently, Mars only has little water perhaps from meteorites or other sources.

Today, scientists know many large Earth-like planets (known as super-Earths) outside the solar system. An Earth-like exoplanet with 3 Earth masses is predicted to have a core-boundary pressure of 500 GPa, while a Mercury-like exoplanet with 5 Earth masses has a similar core-boundary pressure [33], such as Luyten b [34] and Gliese 625 b [35]. For these exoplanets, $Mg_2SiO_5H_2$ should exist as a water reservoir, not only before the core–mantle separation but also throughout their entire history. Therefore, a very different geochemical cycle of water is expected for super-Earths. The likely consequences include: 1) the superionic state of $Mg_2SiO_5H_2$ should contribute to the magnetic field of the planet, in addition to the field produced by convection of the iron core [36]. 2) Mantle-derived water cycle can contain huge amount of water to maintain exposed continents which is necessary for a habitable environment: previous study [37] supported an exoplanet with any mass, as long as its water mass fraction is less than ~0.2%, will maintain exposed continents. Obviously, our new mantle water source of $Mg_2SiO_5H_2$ with 11.4 wt % can increase this limit of water mass fraction of 0.2% by several times to enlarge the field of habitable exoplanet.

In summary, using a comprehensive structure search, we predicted a new stable hydrosilicate $Mg_2SiO_5H_2$. It has two polymorphs, α- and β-$Mg_2SiO_5H_2$, stable at pressures of 262–338 GPa and >338 GPa, respectively. β-$Mg_2SiO_5H_2$ is the first case of Mg in a 9-fold coordination. These new hydrous silicates show much better thermal stability than other known hydrous minerals at megabar pressures and temperatures of thousands of Kelvins. They are suggested to have acted as water reservoirs before and in the early stage of the core–mantle separation in the newborn Earth. In the later parts of the formation of the Earth's core, $Mg_2SiO_5H_2$ was displaced to depths(pressures) where it was no longer stable and released water. This now extinct compound has thus likely played a very important role in shaping the evolution of the Earth, and affecting its physics and chemistry.

This work was supported by NSFC (Grants No. 21803033, No. 12174200, No. 52025026, No.11874224, No. 52090020), Young Elite Scientists Sponsorship Program by Tianjin (Grant No. TJSQNTJ-2018-18), Nature Science Foundation of Tianjin (Grant No. 20JCYBJC01530), and the foundation support from the Laboratory of Computational Physics (No. 6142A05200401) and United Laboratory of High-Pressure Physics Earthquake Science (2020HPPES03). The calculations were performed and supported by Tianhe II in Guangzhou and Supercomputing Center of Nankai University (NKSC). A.R.O. acknowledges funding from the Russian Science Foundation (grant 19-72-30043).